\documentclass[iop]{emulateapj}
\usepackage{apjfonts}
\usepackage{natbib}
\usepackage{amsfonts}
\usepackage{amsmath}
\usepackage[usenames]{color}
\usepackage{ulem}
\usepackage{verbatim}
\usepackage{longtable}
\usepackage{array}
\usepackage{epsfig}
\usepackage{graphicx}
\usepackage{threeparttable}
\newcommand\msun{M$_\odot$}
\newcommand\lsun{L$_\odot$}

\newcommand\pasa{PASA}
\newcommand\rh{R$_{\rm h}$}
\usepackage{epsfig}

\shorttitle{NGC 3628-UCD1}
\shortauthors{Jennings et al.}

\begin{document}

\title{NGC 3628-UCD1: A possible $\omega$~Cen Analog Embedded in a Stellar Stream}

\author{Zachary G. Jennings\altaffilmark{1},
Aaron J. Romanowsky\altaffilmark{1,2},
Jean P. Brodie\altaffilmark{1},
Joachim Janz\altaffilmark{3},
Mark A. Norris\altaffilmark{4},
Duncan A. Forbes\altaffilmark{3},
David Martinez-Delgado\altaffilmark{5},
Martina Fagioli\altaffilmark{3}, 
Samantha J. Penny\altaffilmark{6}
}

\altaffiltext{1}{University of California Observatories, Santa Cruz, CA 95064, USA; zgjennin@ucsc.edu}
\altaffiltext{2}{Department of Physics and Astronomy, San Jos\'{e} State University, One Washington Square,
San Jose, CA, 95192, USA}
\altaffiltext{3}{Centre for Astrophysics \& Supercomputing, Swinburne University, Hawthorn VIC 3122}
\altaffiltext{4}{Max Planck Institut f\"{u}r Astronomie, K\"{o}nigstuhl 17, D-69117 Heidelberg, Germany}
\altaffiltext{5}{Astronomisches Rechen-Institut, Zentrum f\"{u}r Astronomie
der Universit\"{a}t Heidelberg, M\"{o}nchhofstr. 12-14, 69120 Heidelberg,
Germany}
\altaffiltext{6}{Institute of Cosmology and Gravitation, University of Portsmouth, Dennis Sciama Building, Burnaby Road, Portsmouth PO1 3FX, UK}

\begin{abstract}
Using Subaru/Suprime-Cam wide-field imaging and both Keck/ESI and LBT/MODS spectroscopy,
we identify and characterize a compact star cluster, which we term NGC 3628-UCD1, embedded
in a stellar stream around the spiral galaxy NGC 3628.
The size and luminosity of UCD1 are similar to
$\omega$~Cen, the most luminous Milky Way globular cluster,
which has long been suspected to be the stripped remnant of an accreted dwarf galaxy.
The object has a magnitude of $i=19.3$ mag (${\rm L}_{\rm i}=1.4\times10^{6}$~\lsun).
UCD1 is marginally resolved in our ground-based imaging,
with a half-light radius of $\sim10$ pc.
We measure an integrated brightness for the stellar stream of
$i=13.1$ mag, with $(g-i)=1.0$. This would correspond to an accreted dwarf galaxy with an approximate luminosity of 
${\rm L}_i\sim4.1\times10^{8}$~\lsun.
Spectral analysis reveals
that UCD1 has an age of $6.6$ Gyr , $[\rm{Z}/\rm{H}]=-0.75$, and
$[{\alpha}/\rm{Fe}]=-0.10$.
We propose that UCD1 is an example of an $\omega$~Cen-like star cluster possibly forming from
the nucleus of an infalling dwarf galaxy, demonstrating that at least some of the
massive star cluster population may be created through tidal stripping.
\end{abstract}

\section{INTRODUCTION}
Since the discovery of ultra-compact dwarfs (UCDs) a decade and a half ago
\citep{hilker1999,drinkwater2000}, there has been considerable discussion in the
literature regarding their origin. The conversation can be distilled down to a simple question:
are UCDs the largest star clusters, or the smallest galaxies?

The earliest UCDs discovered have \rh
$\sim$20 pc and luminosities $>10^{7}$~\lsun. These objects represent a middle-ground
between globular clusters (GCs), which
have~\rh~of $\sim$3 pc and luminosities of $\sim10^{6}$~\lsun,
and dwarf galaxies, with~\rh~$>100$ pc.
Expanded observational studies have found that UCDs
occupy a sequence with similar luminosity to GCs, but larger \rh~\citep{brodie2011,misgeld2011,norris2014}.
We adopt the definition of UCDs from \citet{brodie2011}: UCDs
are objects with \rh~ranging from $\sim$10 to 100 pc,
and luminosities ${\rm M}_{\rm i} < -8.5$ mag (or L$_{\rm i} \gtrsim 10^{5} $L$_{\odot}$).

The simplistic galaxy vs.~cluster distinction breaks down further.
UCDs could include objects resulting from mergers of globular clusters (e.g.
\citealt{fellhauer2002,kissler-patig2006}), or objects formed primordially in
intense star formation episodes~\citep{murray2009}. Characterizing the UCD population
would have implications for cluster formation physics.
In the galaxy scenario, these objects could
form primordially in association with distinct dark-matter halos,
or they could be the remnant nuclei of larger
galaxies which have undergone tidal stripping during accretion onto larger galaxy halos
(e.g.~\citealt{bekki2001,pfeffer2013}). Understanding the origins of the UCD
population has important implications for understanding sub-halos in a $\Lambda$CDM context.

\begin{figure*}
\includegraphics[width=\textwidth]{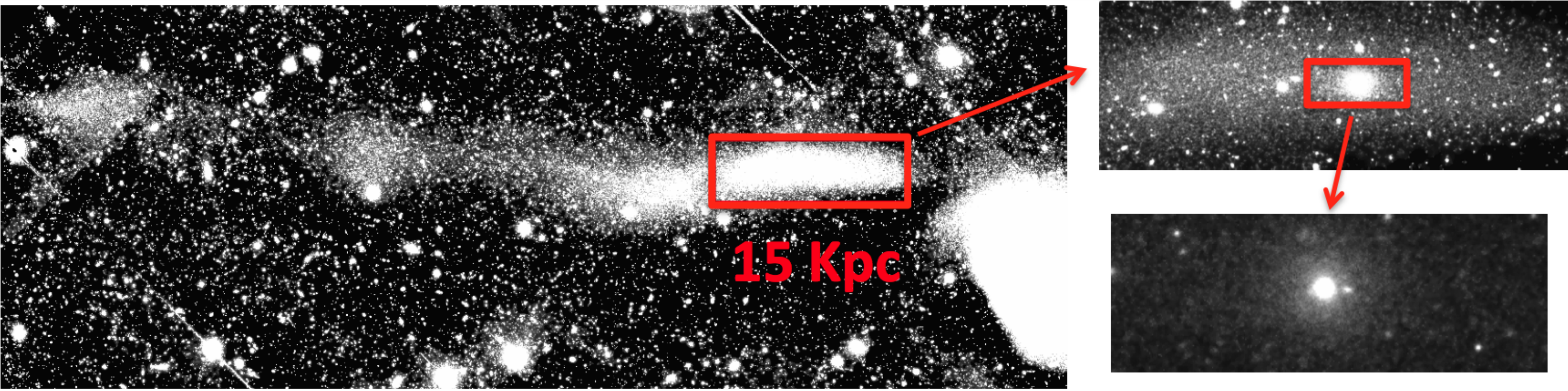}
\caption{Smoothed image of stellar stream next to NGC 3628 from our $i$-band
Subaru/SuprimeCam imaging. The left-most edge of NGC 3628 is visible at the far right of the image.
We highlight the location of UCD1 in zoom-in panels. North is up and east is left.
The limiting surface brightness in the large image
is roughly $\mu_i\sim28.5$ mag arcsecond$^{-2}$. The stretch is modified in each
image. Angular sizes are approximately 26x10 arcmin, 4.5x1.5 arcmin, and 0.9x0.4 arcmin from largest
to smallest scale.}
\end{figure*}

The most massive Milky Way (MW) GC, $\omega$~Cen, is an outlier
among the MW population and may be an example of a sripped dwarf-galaxy nucleus. 
The cluster has a large velocity dispersion (e.g.~\citealt{anderson2010}),
rapid rotation leading to flattening (e.g.~\citealt{merritt1997}),
unusual abundance patterns revealing multiple populations (e.g.~\citealt{king2012}),
and odd orbital properties (e.g.~\citealt{dinescu1999}).
However, definite confirmation of this formation scenario has remained elusive.

In this work, we identify and describe a star cluster,
which we call NGC 3628-UCD1 (hereafter UCD1), embedded in a
stellar stream around the nearby spiral galaxy NGC 3628.
NGC 3628 is an Sb galaxy with ${\rm M}_{\rm V} = -21.37$. It is located in the Leo Triplet, a loose
group with two other large companions, NGC 3623 and NGC 3627. An obvious stellar stream extends
$\sim$140 kpc away from the galaxy, shown in Fig.~1. UCD1 is located
within the plume nearest to NGC 3628. As shown in \S2,
UCD1's size and luminosity are very similar to those of $\omega$~Cen.

The stream itself has been
studied extensively since its first characterization \citep{zwicky1956,kormendy1974}.
The stream contains significant neutral hydrogen \citep{rots1978,haynes1979}.
\citet{chromey1998} identified two blue clumps along the stream and estimated young ages for both.

There are several known examples in the literature of UCDs connected with tidal stripping events.
\citet{norris2011} identified a young UCD around NGC 4546 and argued,
based on the properties of the galaxy,
that it is a result of stripping. \citet{foster2014}
identified an object likely to be the nucleus of the dwarf galaxy forming the "umbrella stream" around NGC 4651.
\citet{mihos2015} identify a nucleus of an ultra diffuse galaxy in Virgo, which they argue is in the process of tidal threshing. 

Using Subaru/Suprime-Cam imaging, we measure photometry and size of both UCD1 and
the full stellar stream.
We propose that UCD1 is an example
of a UCD in formation through tidal stripping.
By measuring the UCD and inferring the properties of the potential progenitor galaxy,
we are able to draw an evolutionary
line in size/luminosity parameter space between the original parent galaxy and the stripped
cluster.
In \S2, we explain our imaging analysis and results, and in \S3, we do the same for our spectroscopy.
We discuss our results in \S4.

NGC 3627 has a
Cepheid distance measurement of 10.6 Mpc \citep{kanbur2003}.
We adopt this value for NGC 3628 and
include an approximate distance uncertainty of $\pm$1 Mpc on distance-dependent properties,
given the potential offset of NGC 3628 from NGC 3627. 

\section{Imaging}

\subsection{Data Reduction}
We imaged NGC 3628 in $r$-band for 425s on 2009 April 20th
with Subaru/Suprime-Cam.
UCD1, a marginally-resolved source in the center of plume just east of NGC 3628,
was first noticed in this pointing.
We subsequently acquired
imaging centered on UCD1 in the $i$-band on 2014 March 3 (425s exposure time)
and in the $g$-band
on 2014 December 19 (1225s exposure time).
Seeing was $\sim0.80^{\prime\prime}$,  $\sim0.70^{\prime\prime}$, and $\sim0.75^{\prime\prime}$
in $g,r,i$ respectively.
We employed a modified version of the SDFRED-2
pipeline\footnote{http://subarutelescope.org/Observing/Instruments/SCam/sdfred/sdfred2.html.en}
to reduce our Suprime-Cam data. AB Zeropoints were calculated
by comparing photometry for bright, unsaturated stars in both the SDSS catalog and the Suprime-Cam
imaging. We used the \citet{schlafly2011} values from NED to correct for Galactic extinction. 

\begin{table}
\begin{center}
\begin{tabular} {c c c}
\multicolumn{3}{c}{UCD1 Properties} \\
Parameter & Value & Uncertainty \\
\hline
R.A. (J2000, from SDSS) & 170.25493 & - \\
Dec (J2000, from SDSS) & 13.60813 & - \\
 & & \\
\hline
\multicolumn{3}{c}{Suprime-Cam Photometry}\\
$g$ & 19.98 mag & 0.05 mag \\
$r$ & 19.57 mag & 0.04 mag \\
$i$ & 19.29 mag & 0.04 mag \\
$(g-r)$ & 0.41 mag & 0.05 mag \\
$(g-i)$ & 0.69 mag & 0.05 mag \\
\rh, $r$-band & 10 pc & 3 pc \\
Ellipticity & 0.9 & - \\
Luminosity & ${\rm L}_{\rm i}=1.4\times10^{6}$~\lsun & $\pm0.2\times10^{6}$~\lsun \\
 & & \\
 \multicolumn{3}{c}{ESI Spectroscopy}\\
Vel. & 815 km s$^{-1}$ & 4 km s$^{-1}$\\
 Vel. Dispersion & $\lesssim 23$ km s$^{-1}$ & - \\
Age & 6.6 Gyr & $^{+1.9}_{-1.5}$ Gyr \\
$[\rm{Z}/\rm{H}]$ & -0.75 & 0.12  \\
$[{\alpha}/\rm{Fe}]$ & -0.10 & 0.08 \\
& & \\
 \multicolumn{3}{c}{MODS Spectroscopy}\\
Age & 6.6 Gyr & $^{+1.4}_{-1.2}$ Gyr \\
$[\rm{Z}/\rm{H}]$ & -0.77 & 0.16  \\
$[{\alpha}/\rm{Fe}]$ & -0.08 & 0.15 \\
& & \\
\hline
& & \\
& & \\
& & \\
\multicolumn{3}{c}{Surface Photometry}\\
Filter & Apparent Mag & Luminosity \\
\hline
\multicolumn{3}{c}{Full Stream}\\
$g$ & 14.15 mag & $(2.7\pm0.5)\times10^{8}$ \lsun\\
$i$ & 13.14 mag & $(4.1\pm0.8)\times10^{8}$ \lsun\\
$(g-i)$ & 1.01 mag & - \\
\multicolumn{3}{c}{Plume Containing UCD1}\\
$g$ & 15.34 mag & $(9.^{+1.8}_{-1.7})\times10^{7}$ \lsun \\
$i$ & 14.19 mag & $(1.6\pm0.3)\times10^{8}$ \lsun\\
$(g-i)$ & 1.15 mag & - \\
\end{tabular}
\end{center}
\end{table}

\subsection{Photometry and Size of UCD1}
We performed aperture photometry of UCD1 using an aperture roughly
twice the size of the FWHM for each image, chosen to maximize the S/N.
Aperture corrections were measured using several bright,
unsaturated stars in the field. Uncertainties in aperture corrections were 0.04 mag in $g$ and 0.03 mag in $r$ and $i$.
Choice of sky subtraction annulus introduced systematic uncertainties of 0.03 in all filters, with no effect on color. 
Our photometry is listed in Table~1. At our assumed distance, the
luminosity of UCD1 is ${\rm L}_{\rm i}=(1.4\pm0.2)\times10^{6}$~\lsun\,with the uncertainty dominated by the distance.
The color measured for UCD1 is dependent on the aperture selected.
When an aperture equal to the FWHM is used, the color is $g-i=0.86$, which
is comparable to $g-i=0.91$ inferred from the stellar populations in the ESI spectrum (see \S3). Note that the seeing FWHM
in our imaging is roughly comparable to the width of the ESI slit used.
Using the max S/N aperture results in bluer colors ($g-i=0.69$).
This effect may be due to contamination by the stream, or could be caused by some gradient
intrinsic to UCD1. Robust determination
of UCD1's color will require more sophisticated stream/source decomposition.
Given the varying quality of our ground-based images,
our current dataset is not well-suited to this task.

UCD1 is marginally resolved in our imaging;
we used \textit{ishape} \citep{larsen1999} to measure~\rh, exploring
Sersic and King profile fits.
\textit{ishape} convolves a model light profile with an empirical PSF and fits it to the source.
We measured the PSF from bright, unsaturated stars in the FOV.
Both Sersic and King profiles feature a parameter to describe the shape of the profile.
When left completely free, the resulting fits featured unphysical values for shape parameters.
However, we found that varying these parameters over a reasonable range
changed \rh~at roughly the 20\% level. 

Across all filters, a model+PSF was always a better fit than the PSF-only model.
As the $r$-band imaging features the best seeing, we adopt our $r$-band fits for UCD1's fiducial \rh.
In general, fits to the $i$-band data tended to be $\sim$10\% smaller, while fits in $g$-band
data tended to be $\sim$10\% larger than $r$-band measured \rh.

We find a size of 7.5 pc for a $c=15$ a King profile and
12.5 pc for an $n=4$ Sersic profile
The differences in residuals are not
large between the two assumed models. We ultimately choose to adopt the average of the two values and
note a $\pm$2 pc systematic uncertainty on this measurement depending on the choice of profile.
We also note the 10\% uncertainty
from our assumed distance, as well as $\sim$20\% scatter for choices of concentration parameter.
Our final $r$-band \rh~estimate for UCD1 is thus $10\pm3$ pc.

\subsection{Surface Photometry of NGC 3628 Stream}
One possible explanation for the stellar stream is the accretion of a dwarf galaxy.
If this scenario is true, then the total luminosity of the NGC 3628
stream offers a useful constraint on the luminosity
of this accreted dwarf. 

The Suprime-Cam
reduction pipeline subtracts a constant sky brightness off the image.
We use SExtractor to model any remaining varying background, adopting a mesh size of 512 pixels. This size was chosen
to be significantly larger than the stream. We inspected the background maps to verify the stream was not being included.
Bright objects were masked in the imaging.
We then use \textit{adaptsmooth} \citep{zibetti2009,zibetti2009b} to
perform adaptive smoothing to a uniform signal to noise ratio (S/N) on the background-subtracted images.
We select a S/N threshold of five (limiting surface brightness of $\mu_i\sim28.5$ mag arcsecond$^{-2}$).

We performed aperture photometry of the stream in this smoothed image using a custom aperture (the same
for all filters) and the \textit{polyphot} task in IRAF,
with pixels below the S/N threshold excluded from the measurement.
The resulting measurements are listed in Table~2.

We measure a total apparent $i$-band magnitude of 13.14. Statistical uncertainties are
small (a few thousandths of a magnitude), but doubling or halving the required signal-to-noise
modifies the final answer by around 0.1 mag. We adopt $\pm0.1$~mag as a rough estimate of the
systematic uncertainty in the measurement, which is smaller than the distance uncertainty ($\pm$0.2 mag).
The total luminosity of the stream is L$_i\simeq(4.1\pm0.8)\times10^{8}$ \lsun.
We emphasize that this estimate could miss additional starlight below our detection threshold or behind
NGC 3628, or could include contamination from faint contaminant point sources, and therefore should
be regarded with caution when interpreted directly as the accreted galaxy luminosity. We measure
$g-i=1.01$ for the color of the full stream.

The plume which contains UCD1 has $i=14.19$
and $g=15.34$, giving $(g-i)=1.15$. The approximate area used for the aperture photometry
is marked in Fig.~1. Using the galaxy stellar population
models of \citet{into2013}, the corresponding $i$-band stellar mass-to-light ratio (M/L$_{*}$) is 2.5--2.9.
This color may be more representative of the stellar population of the progenitor dwarf galaxy, as the
full-stream measurement can be more sensitive to contaminants and choice of aperture.
Both the plume and full stream color and luminosity measurements fall on the red side of galactic scaling relations
from \citet{janz2009}, but they still are broadly within the scatter.

\section{SPECTROSCOPY OF UCD1}

\begin{figure}
\includegraphics[width=0.5\textwidth]{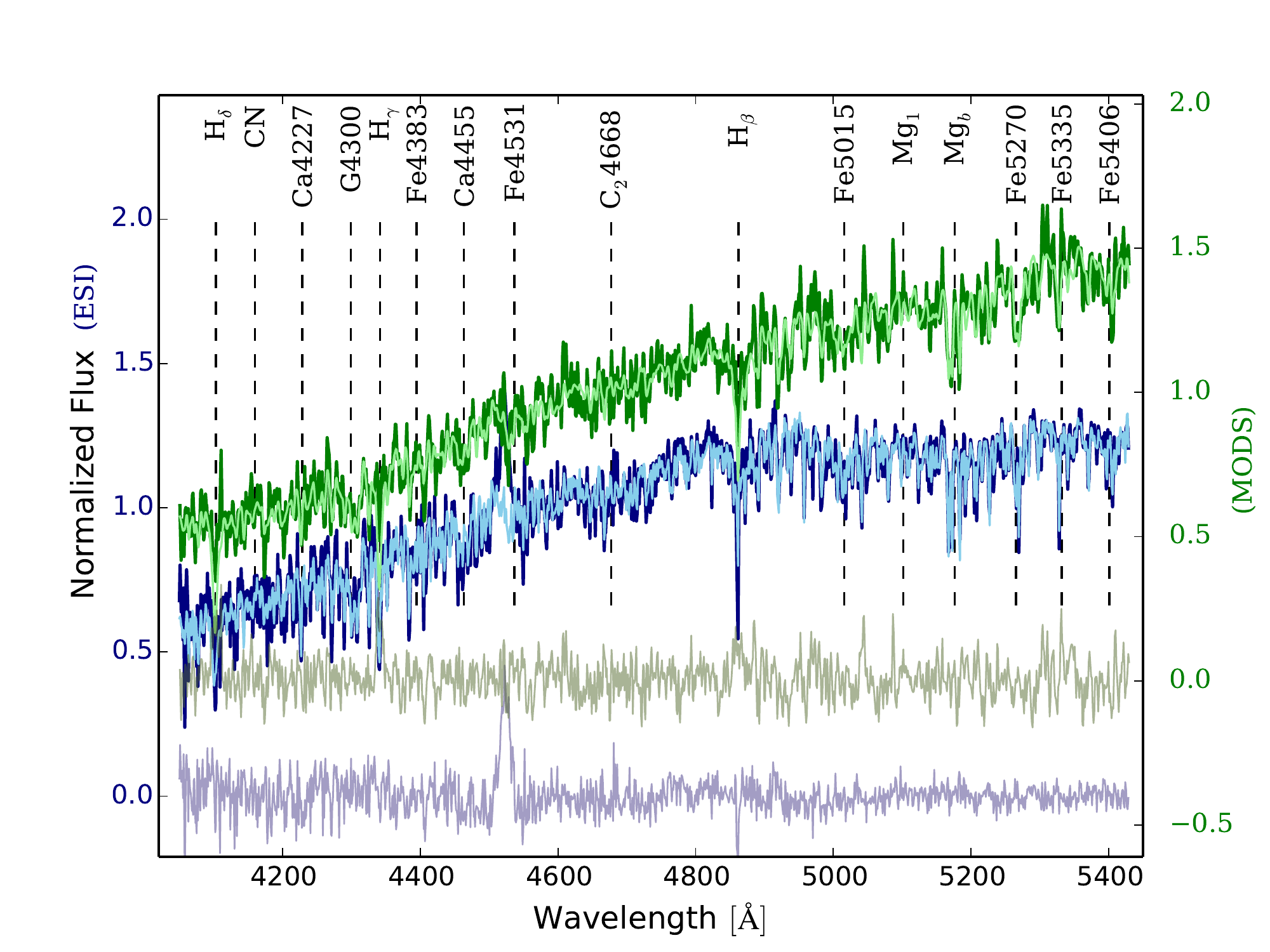}
\caption{Keck/ESI and LBT/MODS spectra of UCD1. ESI data are shown in dark blue, while MODS data are shown
in dark green. Light blue and light green represent model fits to both spectra. Residuals are shown at the bottom.
The MODS scale is offset from the ESI scale.}
\label{spec}
\end{figure}

A 3600s KECK/ESI spectrum of NGC3628 UCD1 was acquired on 2014 March 20th with
the 0.75$^{\prime\prime}$ longslit, with a S/N of $\sim$23 \AA$^{-1}$.
We subsequently acquired a 2400s LBT/MODS spectrum on 2015 June 6th with the 0.80$^{\prime\prime}$
longslit, with a S/N of $\sim$15 \AA$^{-1}$.
We follow the same procedure as \citet{janz2015} to measure stellar populations.
Briefly, we measure all
Lick indices. These are then compared to the single stellar population
models of \citet{thomas2011}, which give the best age, metallicity and alpha element abundance via a
$\chi^2$ minimization process. Poorly fitting lines are excluded from the analysis in an iterative way.
The final fit and residuals to both the observed spectra are shown in Fig.~\ref{spec}. 

For the ESI spectrum, we find an old age of
$6.6^{+1.9}_{-1.5}$ Gyr and a metal-poor population of $[\rm{Z}/\rm{H}]=-0.75\pm0.12$, and alpha-element
abundance of $[{\alpha}/\rm{Fe}]=-0.10\pm0.08$. The MODS spectrum gives consistent results,
with an age of , $6.6^{+1.4}_{-1.2}$ Gyr, $[\rm{Z}/\rm{H}]=-0.77\pm0.16$ and $[{\alpha}/\rm{Fe}]=-0.08\pm0.15$.
Using the relation $[$Fe/H$]=[$Z/H$]-0.94 \times$[${\alpha}$/Fe$]$ \citep{thomas2003},
we find $[{\rm Fe}/{\rm H}]=-0.84$ for the ESI data.

As the ESI spectrum has higher resolution, we use it to measure kinematics (MODS is limited to $\sigma\sim$55 km s$^{-1}$).
We measure a heliocentric velocity of 815$\pm$4 km s$^{-1}$.
While we find a best-fit velocity dispersion of 10.5 km s$^{-1}$, this measurement is significantly
below the resolution of the ESI spectrograph ($\sigma=$23 km s$^{-1}$).
Velocity dispersions so low include unquantified systematics, and it is unclear if meaningful
constraint can be obtained that low (see also \citealt{geha2002,janz2015}). As a result,
we adopt $\sigma=$23 km s$^{-1}$ as the upper limit for the velocity dispersion.

We may make an estimate of the dynamical mass of UCD1 using the expression
$M_{\rm dyn} = C G^{-1} \sigma^{2} R$,
where $R$ is taken to be the half-light radius and
$C$ is the viral coefficient. We follow \citet{forbes2014}
and adopt a value of 6.5, although values between 4 to 7.5 are reasonable.
Given this expression and using our measured size of 10 pc, we estimate $M_{\rm dyn}$ to be
less than $\sim8\times10^{6}$~\msun.

There is a significant offset of $\sim$75 km s$^{-1}$
between the measured heliocentric velocity of UCD1 and the
heliocentric velocity measured from the HI gas in the stream. \citet{nikiel-wroczynski2014} found a gas
velocity of $\sim$890 km s$^{-1}$ in the vicinity of UCD1.
While this could indicate that UCD1 is not associated with the stream,
we find this unlikely. UCD1 appears directly at
the center of the brightest plume in the stream, and displays a blotchy morphology
similar to that in the stream itself. 
UCD1 is the brightest point in the stream,
with the surface brightness of the stream falling off slowly in all directions away from UCD1. It is unclear
that we would expect the velocity of the HI gas to follow that of the stream; the gas obeys
different physics than the stars in the stream (e.g.~ ram pressure stripping),
and so it would not be surprising to see a velocity offset
between the two. UCD1 is blue-shifted by $\sim$30 km s$^{-1}$ compared to NGC 3628,
while the gas is red-shifted by $\sim$45 km s$^{-1}$.

\begin{figure*}
\centering
\begin{minipage}[t]{0.48\textwidth}
\includegraphics[width=\textwidth]{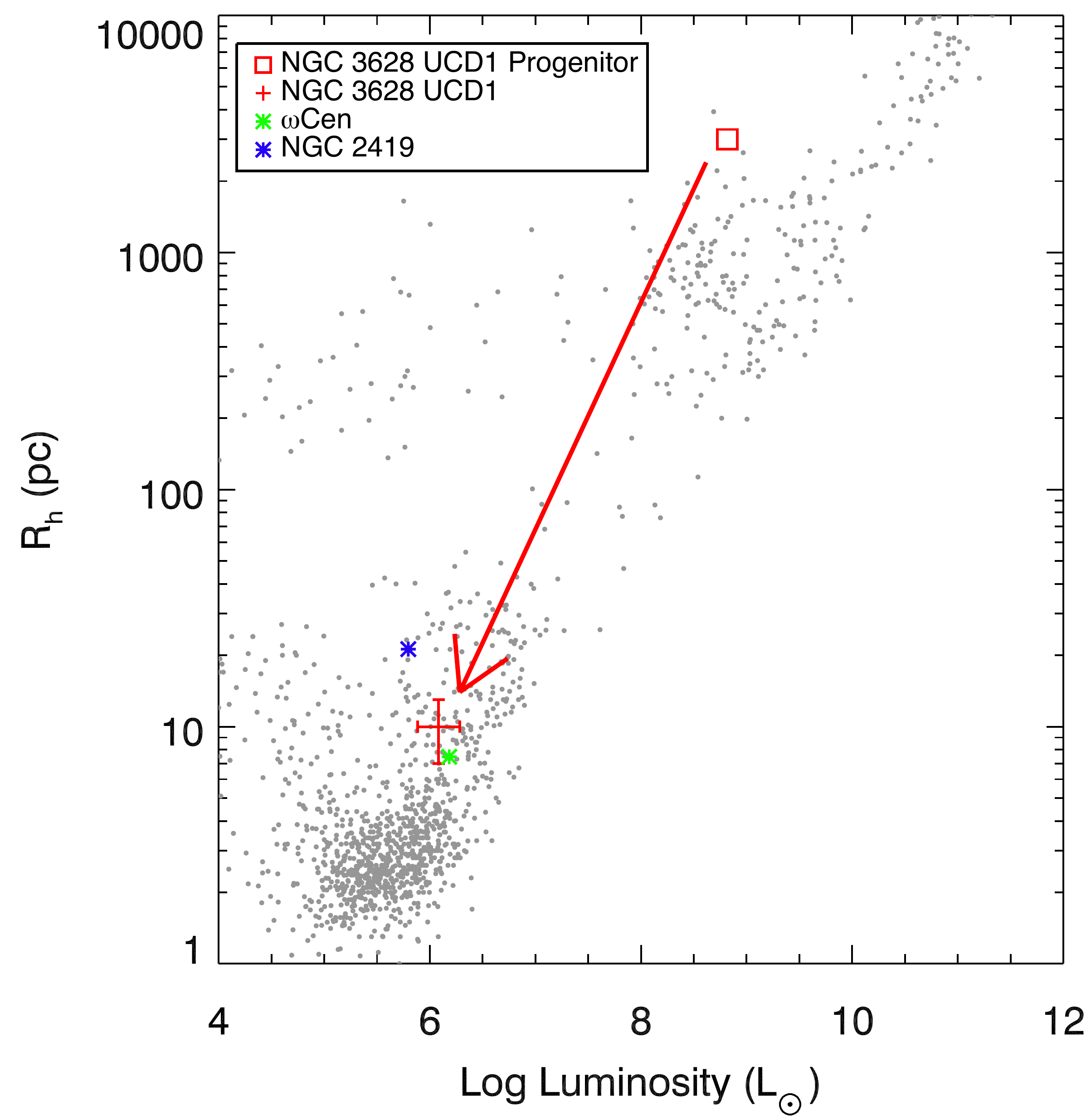}
\end{minipage}
\begin{minipage}[t]{0.48\textwidth}
\includegraphics[width=\textwidth]{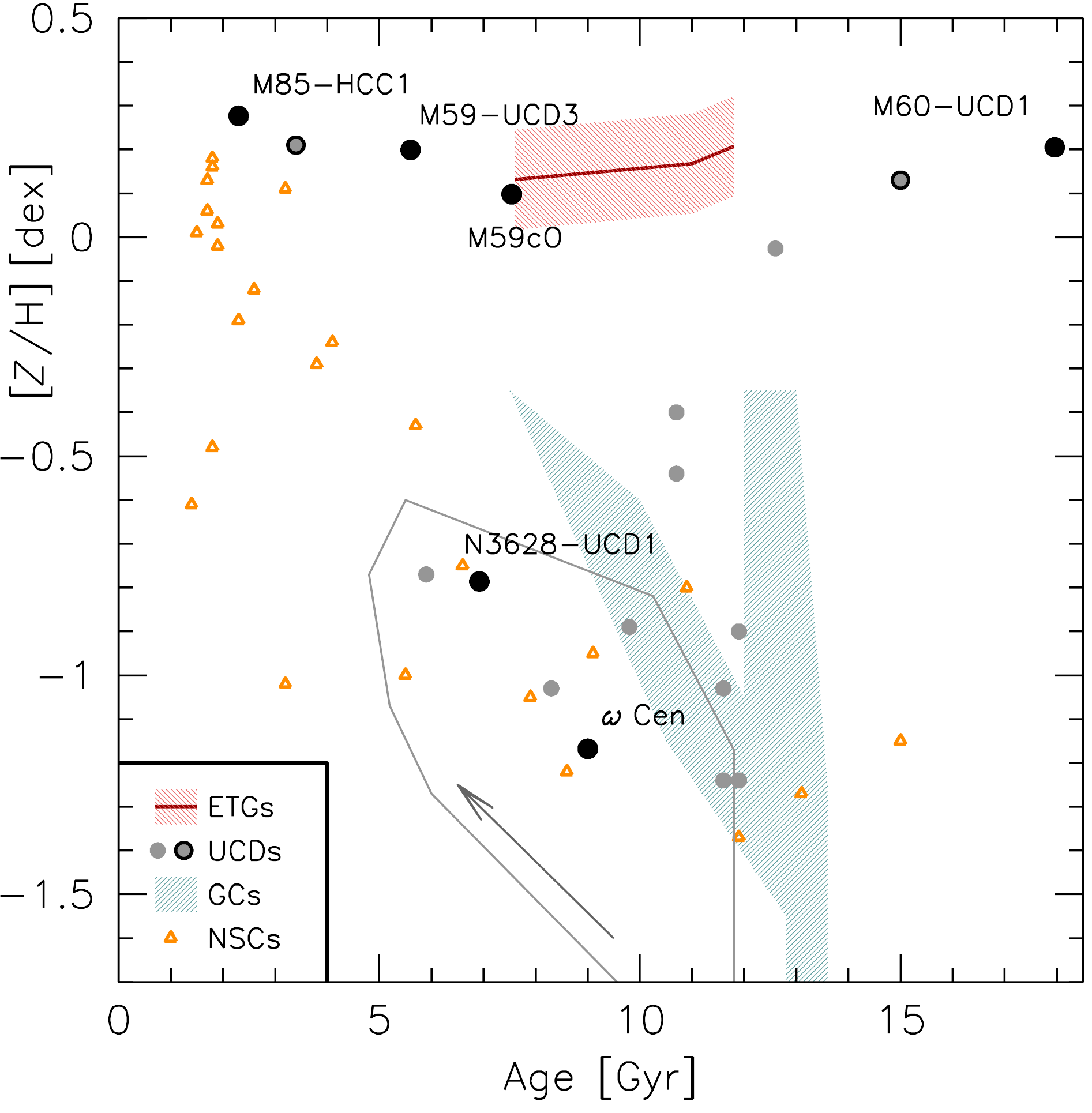}
\end{minipage}
\caption{Left panel: Plot of \rh~(pc) vs.\ $V$-band luminosity (L$_{\odot}$) for
a collection of distance-confirmed dispersion supported stellar systems. Data are from
a compilation begun in \citet{brodie2011},
updated in \citet{brodie2014}.
The location of UCD1 is marked.
We also mark the location of the MW GCs $\omega$~Cen and NGC 2419 for comparison.
We mark the location where the UCD1 progenitor galaxy 
may have originated based on the stellar stream
(note that the box size of this point is arbitrary and not indicative of uncertainty).
Right panel: plot of age (Gyr) vs.\  $[\rm{Z}/\rm{H}]$, modified from \citet{sandoval2015}.
The grey outline corresponds the approximate ranges in both quantities among stars within $\omega$~Cen,
and the arrow indicates the expected direction of evolution \citep{villanova2014}.
UCD1 and $\omega$~Cen are compared to centers of ETGs (red shaded region), 
MW GCs (green shaded region), confirmed UCDs (grey points), and NSCs (orange triangles). See
\citet{pritzl2005,brodie2011,dotter2011,conroy2014}.
Clusters with likely tidal-stripping origins
are also plotted (see \citealt{sandoval2015}.}
\end{figure*}

\section{DISCUSSION AND SUMMARY}
\subsection{The Origin of the Tidal Stream}
In the literature, this stellar stream has often been considered
the result of an interaction between NGC 3628 and another galaxy in the group,
typically NGC 3627. However,
the properties of UCD1 give us cause to consider a potential dwarf galaxy minor-merger
scenario as the source of the stream.

\citet{rots1978}
carried out a restricted 3-body simulation of a tidal interaction between NGC 3627
and NGC 3628. 
The simulation
did a reasonable job of reproducing some features but also had some discrepancies with observations,
such as a smaller velocity difference between the two
large galaxies than was actually observed (a full list of issues is enumerated in \citealt{haynes1979}).
This early simulation also did not take into account other bodies in the system, an important caveat given that the galaxies
share a group environment with NGC 3623 to the south east.

Now that we have identified and characterized UCD1, a successful simulation must be
able to explain the presence of a $\sim10^{6}$~\msun~ compact star cluster with a $\sim$6 Gyr, metal-poor
stellar population.
It is not immediately obvious how such an object could result
simply from the interaction of NGC 3627 and NGC 3628, as we would expect objects resulting from such
an interaction to be younger, metal-rich, and less compact.
It is important to note that, in this work, we are not making any direct comment on the tidal
interaction. We have not yet conducted dynamical modeling of our own, and so we are not in a firm
position to endorse a minor-merger picture for this tidal interaction over the NGC 3628/NGC 3627 interaction
model currently in the literature. The study of this system would benefit greatly
from more sophisticated dynamical
modeling, exploring the possibility of minor mergers and including the dynamical effects of all three
group galaxies. For the moment, we consider the nature of the merger an open question.

\subsection{UCD1 compared to other UCDs}
It is informative to consider the properties of UCD1 in the context of other UCDs. In the left panel of Fig.~3,
we plot the measured half-light radius in pc against the absolute V-band magnitude for a large collection
of distance-confirmed dispersion-supported stellar systems across a wide range of sizes
and luminosities  \citep{brodie2014}.

In the right panel of Fig.~3, we plot the age and metallicity of various star clusters including confirmed
UCDs, MW GCs, nuclear star clusters (NSCs), and early-type galaxy (ETG) centers. 
We also hightlight an assortment of metal-rich UCDs with a likely tidal stripping origin
(e.g.~M60-UCD1, see \citealt{sandoval2015}).

$\omega$~Cen
represents an interesting analogue to UCD1 in these parameters. Its size and luminosity are very
similar to UCD1 and both are low metallicity compared to other clusters
with a likely tidal origin. The Harris 1996 (2010 edition)\nocite{harris1996}
catalog lists values of $\sigma=16.8$
km s$^{-1}$ (e.g. \citealt{mclaughlin2005}), $M_{V}=-10.26$, and \rh$=7.5$ pc (e.g. \citealt{vandenbergh1991}),
all comparable to UCD1 within the uncertainties (note that we
only estimate an upper limit for $\sigma$ for UCD1).
$\omega$~Cen is even more metal-poor than UCD1, $[{\rm Fe}/{\rm H}]=-1.53$ (e.g. \citealt{johnson2009})
vs.~$-0.84$, and, generally speaking, is alpha-enhanced (e.g.~ \citealt{johnson2010}).
Such differences are not unexpected,
since UCD1 is only currently being stripped. Given that $\omega$~Cen was likely accreted several Gyr
ago, UCD1 is naturally expected to have a more extended SF history, enhancing its metallicity and
potentially erasing any alpha-enhancement. In this context, objects like UCD1 may be reasonable modern-day
examples of accretion events which, Gyr ago, would have resulted in $\omega$~Cen-like star clusters. These clusters offer interesting contrasts
in size, luminosity, and metallicity compared to more metal-rich, tidally stripped UCDs.

\subsection{Properties of a Potential Dwarf Galaxy Progenitor}
In this section, we consider the implications if UCD1 did indeed result from a minor merger.
Using the stellar stream luminosity to estimate the stellar mass in the stream is straightforward. As discussed
in \S2.3, for a stream of this color, a $M/L_{*}\sim2.7$ is reasonable.
Given the measured luminosity of $(4.1\pm0.8)\times10^{8}$ \lsun, we estimate a stellar mass
of $\sim1.1\times10^{9}~$\msun. We take this value as a rough estimate of the stellar mass of
a potential accreted dwarf galaxy, also noting neglected systematics from the mass-to-light estimate.
This mass is comparable to the spheroid mass estimated in \citet{norris2011} for the progenitor
of the confirmed stripped nucleus NGC~4546 UCD1 $(3.4^{+1.2}_{-1.5}\times10^9$~\msun).
Following the specific frequencies of \citet{peng2008}, such an accretion event would have contributed
$\sim$9 GCs to the halo of NGC 3628, which could possibly be found as a discrete GC population
in phase space. 

To estimate a width for the stream, we use a region to the east of UCD1 $\sim$100 arcseconds wide and sum up
the light horizontally along the direction of the stream,.
We identify the peak of the light distribution as the stream center, and mark the outer end of the stream as the point
where the light profile approximately flattens. Finally, we define the distance from the center that contains half
the total integrated light of the region as the half-light width of the stream. We find a value of $\sim$ 60$^{\prime\prime}$,
or $\sim$3000 pc. We then adopt this half-light width as a proxy for the \rh~of the accreted parent galaxy. 
In Fig.~3, we plot the estimated location of this
dwarf galaxy in the same parameter space. We emphasize that
the connection between stream-width and progenitor size is unclear, and will at the very least
depend on viewing angles and orbital phase.

If a minor-merger picture is correct, it is natural to compare UCD1 to nuclear star clusters (NSCs). NSCs typically
have effective radii of a few to tens of parsecs, and luminosities from a few times $10^5$~\lsun~to~$\sim10^8$~\lsun~
\citep{georgiev2014}. Early-type galaxies in particular may have more compact NSCs;
\citet{cote2006} found a median \rh~of 4.2 pc for prominent NSCs in early-type galaxies.
The precise effects of the stripping process on NSC \rh~are complicated and depend on a wide range of orbital parameters
and initial conditions, but for most scenarios, we wouldn't expect drastic changes in cluster \rh~\citep{pfeffer2013}.
\citet{bianchini2015} found that, as galaxies are stripped, central clusters expand to reach the sizes they would
have in isolation, i.e.~similar to GC/UCD-sized objects. A late-type nucleated galaxy could be a possible source,
though presumably the bulge would need to be red enough to explain the colors seen in UCD1's plume.

For UCD1, if the properties remain roughly comparable, then the
stripping process has, at the order of magnitude level,
resulted in a decrease of roughly 100 in size and 1000 in luminosity. Similar factors of stripping have been seen
in simulations of dwarf-elliptical galaxies \citep{pfeffer2013}, although the precise amounts of mass loss depend on
the orbital parameters. In any case, we find the properties of UCD1 consistent with the cluster
being the final result of a minor merger, offering support that some portion of the large star cluster
population may be created through the tidal stripping of dwarf galaxies.

\acknowledgements
We thank the anonymous referee for comments which greatly improved the paper.
We thank J. Strader, J. Choi, H. Baumgardt, and J. Pfeffer for useful discussions in the course of this work. ZGJ is supported in
part by an NSF Graduate Research Fellowship. This work was supported by NSF grants AST-1109878, AST-1211995,
AST-1515084, and AST-1518294. D.A.F. and J.J.
thank the ARC for financial support via DP130100388.The LBT is an international collaboration among institutions in the 
United States, Italy and Germany. LBT Corporation partners are: 
The University of Arizona on behalf of the Arizona university system; 
Istituto Nazionale di Astrofisica, Italy; LBT Beteiligungsgesellschaft, 
Germany, representing the Max-Planck Society, the Astrophysical Institute 
Potsdam, and Heidelberg University; The Ohio State University, and The 
Research Corporation, on behalf of The University of Notre Dame, 
University of Minnesota and University of Virginia.
Based in part on data collected at Subaru Telescope, which is operated by the National Astronomical Observatory of Japan.
Some of the data presented herein were obtained at the W.M. Keck Observatory, which is operated as a scientific partnership among the California Institute of Technology, the University of California and the National Aeronautics and Space Administration. The Observatory was made possible by the generous financial support of the W.M. Keck Foundation.
The authors wish to recognize and acknowledge the very significant cultural role and reverence that the summit of Mauna Kea has always had within the indigenous Hawaiian community.  We are most fortunate to have the opportunity to conduct observations from this mountain.


\end{document}